\documentclass[aps,pre,groupedaddress,twocolumn,10pt]{revtex4-1}
\usepackage{graphicx}
\usepackage{epstopdf}
\usepackage[english]{babel}
\usepackage[T1]{fontenc}
\usepackage{verbatim}
\usepackage{float}

\usepackage{makecell}
\usepackage{color}
\usepackage{multirow}
\usepackage{booktabs}
\usepackage{amsmath}
\usepackage{amssymb}
\usepackage{amsthm}
\usepackage{bbm}
\usepackage{hyperref}
\usepackage[T1]{fontenc}
\usepackage[usenames,dvipsnames]{xcolor}
\hypersetup{colorlinks=true, linkcolor=blue, urlcolor=blue!50!black, citecolor=blue!50!black}%BrickRed

\usepackage[normalem]{ulem}

\renewcommand{\imath}[0]{\mathsf{i}}

\usepackage[nodayofweek,level]{datetime}

\begin{document}

\title{Dynamics of droplets driven by electrowetting}

\author{Ke Xiao}
\email{xiaoke@ucas.ac.cn}
\affiliation{Department of Physics, College of Physical Science and Technology, Xiamen University, Xiamen 361005, People's Republic of China}
\affiliation{Wenzhou Institute, University of Chinese Academy of Sciences, Wenzhou 325016, People's Republic of China}
\author{Chen-Xu Wu}
\email{cxwu@xmu.edu.cn}
\affiliation{Department of Physics, College of Physical Science and Technology, Xiamen University, Xiamen 361005, People's Republic of China}

\begin{abstract}
Even though electrowetting-on-dielectric (EWOD) is a useful strategy in a wide array of biological and engineering processes with numerous droplet-manipulation applications, there is still a lack of complete theoretical interpretation on the dynamics of electrowetting. In this paper, we present an effective theoretical model and use Onsager variational principle to successfully derive the governing equations of non-equilibrium electrowetting dynamics for a droplet in both overdamped and underdamped regimes. It is found that the spreading and retraction dynamics of a droplet on EWOD substrates can be fairly well captured in the overdamped regime. By varying liquid viscosity, droplet size, and applied voltage, we confirm that the transient dynamics of EW can be characterized by a timescale independent of liquid viscosity, droplet size and applied voltage. Our model provides a complete fundamental explanation of EW-driven spreading dynamics, which is important for a wide range of applications, from self-cleaning to novel optical and digital microfluidic devices.

\end{abstract}
\date{\today}

\maketitle

%%%MAIN TEXT%%%%
\section{INTRODUCTION}
Electrowetting-on-dielectric (EWOD), a phenomenon referring to the effect of electric fields on the wetting of a droplet on a dielectric layer-covered electrode surface, is an important and versatile technique for droplet manipulation~\cite{F.Mugele2005}. So far it has attracted considerable attention due to its significance in fundamental scientific understanding and numerous technological applications, such as novel digital microfluidic devices~\cite{K.Choi2012,S.JunLee2012,P.G.Zhu2017,J.Hong2015}, fast response displays~\cite{R.A.Hayes2003}, fast optical imaging~\cite{C.L.Hao2014}, optical devices~\cite{S.Kuiper2004,J.Heikenfeld2005}, inkjet/soft printing~\cite{T.Boland2006,L.B.Zhang2015}, self-cleaning~\cite{K.M.Wisdom2013}, anti-icing~\cite{Q.L.Zhang2013,J.B.Boreyko2009}, and inducing droplet detachment~\cite{K.Xiao2021} and wetting transition~\cite{K.Xiao202109}.
To improve the performance of such applications, a fundamental understanding of drop dynamics driven by EW is crucial~\cite{F.Mugele2009}. It has been found that this spreading motion of a droplet on a solid substrate is initiated by the electrical force concentrated near the three-phase contact line (TCL), of which the dynamics is determined by the balance between the driving electrical force, capillary force, and resistance forces (i.e., contact line friction). Earlier studies have revealed that the EW-driven spreading dynamics can be typically categorized into two main regimes, namely the overdamped and the underdamped regimes~\cite{Q.Vo2018SP,Q.Vo2018PRE}. In the overdamped regime dynamic behaviors are dominated by viscous effect~\cite{T.D.Blake1969,L.H.Tanner1979,P.K.Mondal2015,Q.Vo2018SP,Q.Vo2018PRE}, whereas in the underdamped regime, the droplet inertia becomes dominant~\cite{Q.Vo2018PRE,Q.Vo2018SP,J.C.Bird2008,J.Hong2013}.

During the past decade, numerous efforts of experiments~\cite{Q.Vo2018PRE,Q.Vo2018SP,J.Hong2013,M.Marinescu2010,H.Li2013}, theoretical modelings~\cite{J.M.Oh2008,J.M.Oh2010}, and numerical simulations~\cite{S.R.Annapragada2011,Q.Zhao2021} have been devoted to getting a better understanding of the spreading dynamics triggered by EW effect. For example, the dynamic EW and dewetting of ionic liquids have been investigated with high-speed video microscopy, and the experimental measurements have shown that the base area of the droplet varies exponentially during both the EW and retraction processes. The linear dependence of dynamic contact angle on the speed of contact line expansion was examined using the hydrodynamic and molecular-kinetic models~\cite{H.Li2013}. Using the domain perturbation method, J. M. Oh \textit{et al.}~\cite{J.M.Oh2010} analyzed the unsteady motion and the shape evolution of a sessile drop actuated by EW and reached a qualitative agreement between their analytical results and experimental ones, validating their theoretical model. By studying the dependence of spreading dynamics on drop size and viscosity under various voltages, it has been found that there exists a critical viscosity at which the spreading pattern changes from an underdamped one to an overdamped one.
More recently, Q. Vo \textit{et al.} systematically investigated the dynamics of EW droplets, derived the critical viscosity at which the transition occurs, and revealed its subtle and often hidden dependence on the EW dynamics~\cite{Q.Vo2018SP}. They also reported that a transient timescale can be used to characterize both the spreading and retracting dynamics~\cite{Q.Vo2018PRE}. In addition, the volume of fluid method~\cite{S.R.Annapragada2011} and finite element method~\cite{Q.Zhao2021}, in which the fluid dynamics is modeled by the Navier-Stokes equations, were also employed to study the static and the dynamic EW problems.

Despite the fact that the dynamics of droplets induced by EW has been extensively studied via either experimental or theoretical approaches, a systematic analytical understanding of the underlying subtle mechanism, in particular, the governing kinetic equations of the non-equilibrium EW dynamics and the dependence of transient dynamics in EW on liquid viscosity, droplet size and applied voltage, remain elusive.

In this paper, we establish a theoretical model to systematically study the EW-driven dynamics of a droplet on a dielectric layer-covered electrode substrate. The numerical results predicted by the present model are compared with experimental ones. To explore the transient dynamics in both overdamped and underdamped regime, we further investigate how the viscosity and the size of a droplet, and applied voltage effect on EW dynamics. We expect this work can offer some helpful implications for the design of EWOD-based devices when physical properties of droplet (e.g., viscosity and droplet size) and applied voltage are varied to meet their requirements.

\section{THEORETICAL MODELING}
\begin{figure}[htp]
%\centerline{\includegraphics[width=1.0\textwidth,keepaspectratio]{figure2}}
  \includegraphics[width=\linewidth,keepaspectratio]{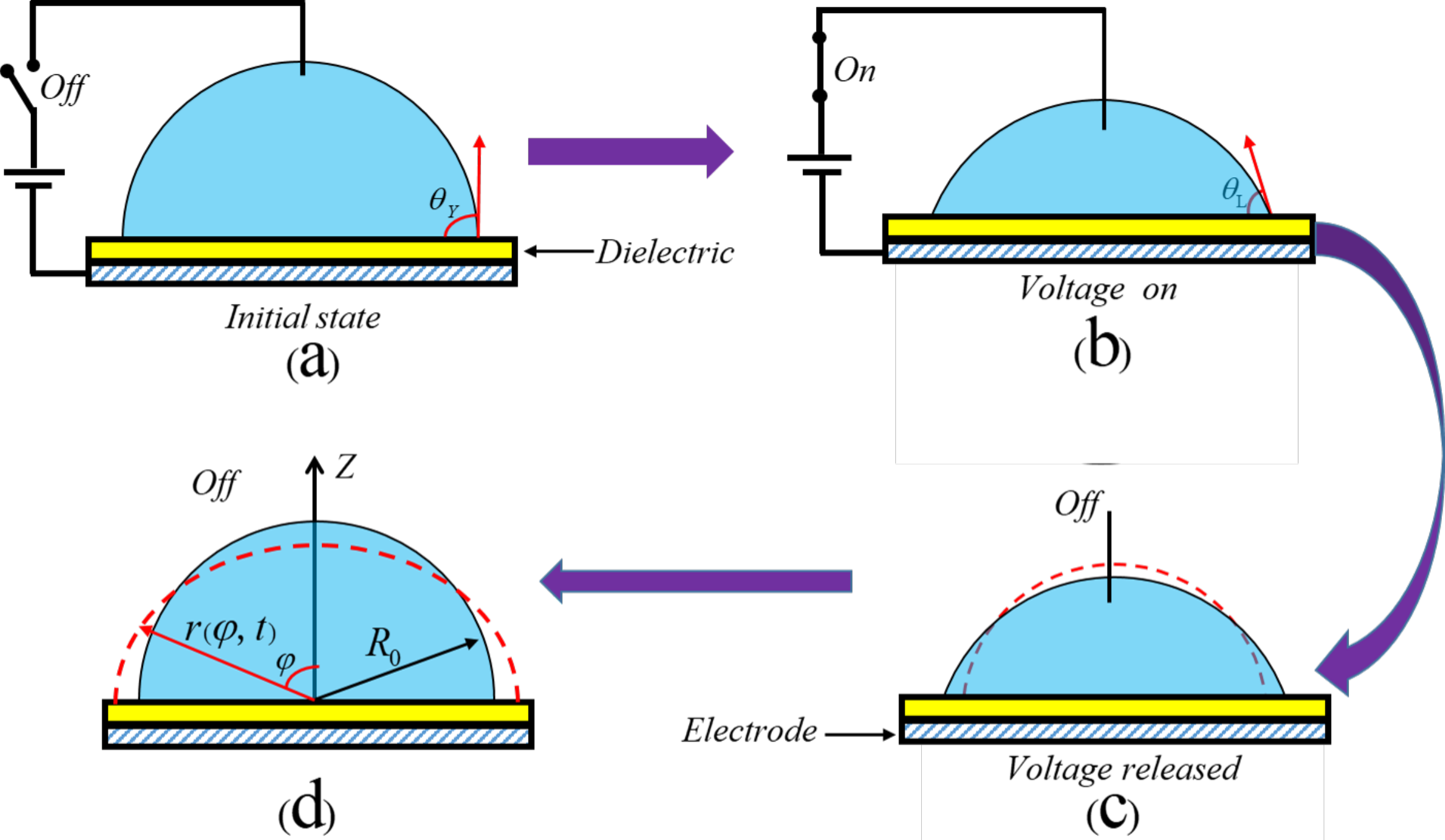}
  \caption{(Color online) Schematic pictures of a droplet on a substrate in equilibrium state with contact angle (a) $\theta_{\rm Y}$ in the absence of electric voltage, (b) $\theta_{\rm e}$ in the presence of electric voltage. (c) Recoiling stage and (d) the equilibrium state of the droplet in the absence of electric voltage. \label{SpredingandRecoiling}}
\end{figure}

We begin our investigation by considering an aqueous droplet placed on a substrate consisting of an insulating surface layer (thickness $d$) on top (yellow) and an electrode underneath, as shown in Fig.~\ref{SpredingandRecoiling}. Both the droplet and the electrode are immersed in an oil fluid (e.g., the surrounding environment is a pool of silicone oil) with viscosity $\mu_0$.
The apparent contact angle $\theta_{\rm Y}$ (see Fig.~\ref{SpredingandRecoiling}(a)) at the equilibrium state satisfies
Young's equation, $\gamma_{\rm sm}-\gamma_{\rm ls}=\gamma_{\rm lm}{\rm cos}~\theta_{\rm Y}$, where $\gamma_{\rm sm}$, $\gamma_{\rm ls}$, and $\gamma_{\rm lm}$ are interfacial tensions of the solid-medium, liquid-solid, and liquid-medium interfaces, respectively. When an external voltage $U$ is applied between a droplet and a flat substrate, the accumulation of free charges near
the electrode causes a reduction of the local liquid-solid surface tension $\gamma_{\rm ls}$ and subsequently induces the spreading of the droplet, yielding another equilibrium state with an apparent contact angle $\theta_{\rm L}$ (see Fig.~\ref{SpredingandRecoiling}(b)) described by the well-known Young-Lippmann equation~\cite{F.Mugele2005,G.Lippmann1875} ${\rm cos}~\theta_{\rm L}={\rm cos}~\theta_{\rm Y}+\eta$, with $\eta=\varepsilon_0\varepsilon U^2/(2d\gamma_{\rm lm})$, where $\eta$, $\varepsilon_0$, and $\varepsilon$ are the dimensionless EW number, the dielectric permittivity in vacuum and the relative dielectric constant, respectively. Here the reduced liquid-solid surface tension can be replaced by an effective one $\gamma_{\rm ls}^{\rm eff}=\gamma_{\rm ls}-\eta\gamma_{\rm lm}$.
Once the electric voltage is switched off, the droplet will keep its current shape as shown in Fig.~\ref{SpredingandRecoiling}(c) as its transient state. However, due to a sudden increase in the local liquid-solid surface tension owing to the short discharge timescale of the droplet-electrode capacitor much faster than the relaxation time of the droplet~\cite{A.Cavalli2016,K.X.Zhang2019}, the restoration
of liquid-solid interfacial tension leads to a recovery effect. Subsequently, the droplet will
undergo a recoiling stage and reach the final equilibrium configuration (see Fig.~\ref{SpredingandRecoiling}(d)), during which a flow will be induced, causing a viscous dissipation typically comprising contributions in the bulk droplet, near the substrate, and at the
vicinity of the contact line~\cite{Q.Vo2018SP,Q.Vo2018PRE,Q.Vo2019}. On one hand, based on the balance of surface tension and electrical force, the static shape of droplets under EW actuation is well understood. On the other hand, the transient droplet shape behaviors, and the transient dynamics of EW droplets are also studied experimentally and theoretically~\cite{J.M.Oh2010,Q.Vo2018SP,Q.Vo2018PRE}. However, there is still a lack of a universal theoretical model to interpret the dynamics of EW droplets. Therefore, the primary aim of this paper is to develop a unified theoretical model that enables us to study the the dynamics of EW droplets. In the next two subsection, a new theoretic model based on Onsager variational principle is established to derive the kinetic equations and shape mode equations that governs the non-equilibrium EW dynamics for the overdamped and the underdamped regimes, respectively. We assume that the volume of the droplet is constant and the gravitational effect is neglected.

\subsection{Overdamped regime}
In the presence of an electric field, the free energy is written as a sum of the interfacial energy
\begin{align}
F=\gamma_{\rm lm}A_{\rm lm}+\gamma_{\rm ls}^{\rm eff}A_{\rm ls}+\gamma_{\rm sm}A_{\rm sm},
\label{GeneralFreeEnergy}
\end{align}
where $A_{\rm lm}$, $A_{\rm ls}$, and $A_{\rm sm}$ are the areas of the liquid-medium, solid-medium, and liquid-solid interfaces, respectively. Let $A_{\rm t}=A_{\rm sm}+A_{\rm ls}$, and with a consideration of the Young-Lippmann equation, Eq.~(\ref{GeneralFreeEnergy}) can be converted to
\begin{align}
F=\gamma_{\rm lm}A_{\rm lm}-\gamma_{\rm lm}A_{\rm ls}\cos\theta_{\rm L}+\gamma_{\rm sm}A_{\rm t}.\label{GFE:V/=0}
\end{align}
Similarly, the general from of the free energy in the absence of an electric field reads
\begin{align}
F=\gamma_{\rm lm}A_{\rm lm}-\gamma_{\rm lm}A_{\rm ls}\cos\theta_{\rm Y}+\gamma_{\rm sm}A_{\rm t}.\label{GFE:V=0}
\end{align}
In the overdamped regime, the shape of the droplet is treated as a sphere due to the dynamic behaviors dominated by viscous effect and the negligible inertial effect. Thereby, for the spreading stage, the free energy of the system can be calculated as
\begin{align}
F_{\rm s}^{\rm o}=\gamma_{\rm lm}\pi R_{\rm b}^2\biggl[\frac{2}{1+{\rm cos}~\theta}-{\rm cos}~\theta_{\rm L} \biggr],\label{GFE:V}
\end{align}
where $R_{\rm b}$, $\theta$, and $\theta_{\rm L}$ represent the base radius, the dynamic contact angle, and the contact angle at equilibrium, respectively. A time derivative of such a free energy leads to
\begin{align}
\dot{F}_{\rm s}^{\rm o}=&\gamma_{\rm lm}2\pi R_{\rm b}\biggl[\dot{R}_{\rm b}(t)\biggl(\frac{2}{1+{\rm cos}~\theta}-{\rm cos}~\theta_{\rm L}\biggr) \notag\\
&+\frac{R_{\rm b}{\rm sin}~\theta}{(1+{\rm cos}~\theta)^2}\dot{\theta}(t) \biggr],\label{rateofGFE}
\end{align}
where the dots denote a derivative with respect to time $t$.
According to $dV(t)/dt=0$, we can obtain
\begin{align}
\dot{\theta}(t)=-\frac{(2+{\rm cos}~\theta){\rm sin}~\theta}{R_{\rm b}}\dot{R}_{\rm b}(t).\label{volumeconstant}
\end{align}
Given the equation for $\dot{\theta}(t)$, one needs one more equation for $\dot{R}_{\rm b}(t)$ to describe the spreading dynamics of EW droplet in the overdamped regime. To get this, we use Onsager principle~\cite{M.Doi2011,M.Doi2013,M.Doi2015}, and determine the evolution
of the system by the minimum of the Rayleighian:
\begin{align}
\Re=\dot{F}+\Phi,\label{Rayleighian}
\end{align}
where $\dot{F}$ is the time derivative of the free energy of the system given by Eq.~(\ref{rateofGFE}), and $\Phi$ is the energy dissipation function. Strictly speaking, the total viscous dissipation is the sum of bulk dissipation and contact line dissipation, However, as the contribution at the vicinity of the contact line dominates the
whole dissipation process in the system~\cite{Q.Vo2019}, we merely consider the
contribution made by the contact line. The dissipation rate for it is expressed as
\begin{align}
\Phi_{\rm vis}^{\rm ct,o}=\pi\lambda R_{\rm b}\dot{R}_{\rm b}^2,\label{dissipation}
\end{align}
where $\lambda$ is the friction coefficient at contact line. Here, the friction coefficient, according to Refs.~\cite{Q.Vo2018SP,Q.Vo2018PRE,Q.Vo2019}, can be given by $\lambda=C(\mu\mu_0)^{1/2}$, where $\mu$ is the droplet viscosity, $\mu_0$ is the viscosity of the surrounding medium of the droplet, and $C$ is a constant depending on roughness and chemical
properties of the surface~\cite{A.Carlson2012}.
Substituting Eqs.~(\ref{rateofGFE}) and (\ref{dissipation}) into Eq.~(\ref{Rayleighian}) and combining Eq.~(\ref{volumeconstant}), leads to
\begin{align}
\Re_{\rm o}=2\pi\gamma_{\rm lm} R_{\rm b}\dot{R}_{\rm b}({\rm cos}~\theta-{\rm cos}~\theta_{\rm L})+\pi\lambda R_{\rm b}\dot{R}_{\rm b}^2.\label{Re}
\end{align}
Onsager principle tells us that $\dot{R}_{\rm b}(t)$ is determined by the
condition $\partial\Re_{\rm o}/\partial \dot{R}_{\rm b}$=0, which gives the the evolution equation
\begin{align}
\dot{R}_{\rm b}=\frac{\gamma_{\rm lm}}{\lambda}({\rm cos}~\theta_{\rm L}-{\rm cos}~\theta).\label{SpreadingkineticEq}
\end{align}
Similarly, in the absence of an electric field, the kinetic equation for the retracting stage can be also expressed as
\begin{align}
\dot{R}_{\rm b}=\frac{\gamma_{\rm lm}}{2\lambda}({\rm cos}~\theta_{\rm Y}-{\rm cos}~\theta).\label{RetractingkineticEq}
\end{align}
Here the retracting motion is purely driven by capillarity.

Therefore, the non-equilibrium spreading dynamics of EW droplets in the overdamped regime is governed by kinetic Eqs.~(\ref{volumeconstant}) and~(\ref{SpreadingkineticEq}), while its recoiling process in the absence of an electric field is ruled by kinetic Eqs.~(\ref{volumeconstant}) and~(\ref{RetractingkineticEq}).

\subsection{Underdamped regime}
Experimentally, it has been reported that low droplet viscosity will lead to an underdamped dynamic features during the droplet spreading process~\cite{J.Hong2013,Q.Vo2018SP}.
As a consequence, the shape of the droplet is no longer spherical in the underdamped regime. To describe the dynamics of EW droplets theoretically, we decompose the shape of an axisymmetric drop into Legendre polynomials with coefficients $c_{2n}$, i.e.
\begin{align}
r(\varphi,t)=R_0+\sum_{n=1}^\infty c_{2n}(t)P_{2n}({\rm cos}\varphi),\label{dropletshape}
\end{align}
where $R_0$, $c_{2n}(t)$, $P_{2n}$, and $\varphi$ are the effective radius of the droplet, the time dependent amplitude of a shape mode, Legendre polynomials, and the polar angle, respectively, as shown in Fig.~\ref{SpredingandRecoiling}(d).
It is worth to note that the rotational symmetry and a $90^\circ$ average contact angle are assumed.

In the situation of low-viscosity flow, the velocity flow is irrotational with a velocity $\textbf{\textit{v}}=\nabla\psi$, where the flow potential $\psi$ satisfies $\Delta\psi=0$. Here we assume the shape of the droplet is symmetric around azimuthal angle, so that the general solution of the Laplace equation in spherical coordinates reads as $\psi(r,\varphi,t)=\sum_{l=0}^\infty [a_l(t)/r^{l+1}]P_l({\rm cos}\varphi)$, where $a_l(t)$ are time dependent coefficients. The boundary condition requires that the perpendicular velocity of the droplet must vanish on the solid substrate ($\varphi=\pi/2$), i.e., $\textit{v}_\varphi=-\partial\psi/r\partial\varphi=0$. To satisfy this condition we should have $l=2n$ and the general solution can be rewritten as $\psi(r,\varphi,t)=\sum_{l=0}^\infty [c_{2n}(t)/r^{2n+1}]P_{2n}({\rm cos}\varphi)$. At the surface of the droplet ($r=R_0$), the radial velocity of the droplet satisfies the boundary conditions $\textit{v}_r=\partial\psi/\partial r=\dot{r}(\varphi,t)$. Then the flow potential can be expressed as
\begin{align}
\psi=(r,\varphi,t)=-\sum_{n=0}^\infty \frac{R_0^{2n+2}}{(2n+1)r^{2n+1}}\dot{c}_{2n}(t)P_{2n}({\rm cos}\varphi).\label{potentialflow}
\end{align}
Similar to the previous subsection, the free energy can be calculated as
\begin{align}
F_{\rm s}^{\rm u}=&\gamma_{\rm lm}\biggl[2\pi R_{0}^2 + 4\pi R_{0}c_0(t) +2\pi \sum_{n=1}^{\infty}\frac{n(2n+1)+1}{4n+1}c_{2n}^2(t) \biggr] \notag\\
&-\gamma_{\rm lm}\biggl[\pi R_{0}^2 + 2\pi R_{0}\sum_{n=1}^{\infty}c_{2n}(t)P_{2n}(0) \biggr]{\rm cos}~\theta_{\rm L}.\label{GFE:UD}
\end{align}
The kinetic energy due to the flow of the fluid is given by
\begin{align}
T=\pi \rho R_0^3 \sum_{n=1}^{\infty}\frac{\dot{c}_{2n}^2(t)}{(2n+1)(4n+1)}.\label{kineticEnergy}
\end{align}
Given these, the free energy change rate $\dot{E}=\dot{F}_{\rm s}^{\rm ud}+\dot{T}$ then is written as
\begin{align}
\dot{E}=&4\pi \gamma_{\rm lm}R_{0}\dot{c}_{0}(t)+4\pi \gamma_{\rm lm}\sum_{n=1}^{\infty}\frac{n(2n+1)+1}{4n+1}c_{2n}(t)\dot{c}_{2n}(t) \notag\\
&-2\pi R_{0}\gamma_{\rm lm}{\rm cos}\theta_{\rm L}\sum_{n=1}^{\infty}\dot{c}_{2n}(t)P_{2n}(0) \notag\\
&+ 2\pi \rho R_0^3 \sum_{n=1}^{\infty}\frac{\dot{c}_{2n}(t)\ddot{c}_{2n}(t)}{(2n+1)(4n+1)}.\label{energychangerate}
\end{align}
Under the volume conservation of the droplet $dV/dt=0$, we get
\begin{align}
\dot{c}_0(t)=-\frac{2}{R_0}\sum_{n=1}^{\infty}\frac{c_n(t)\dot{c}_n(t)}{4n+1}.\label{c0}
\end{align}
Substituting $\dot{c}_0(t)$ into Eq.~(\ref{energychangerate}), one obtains
\begin{align}
\dot{E}=&4\pi \gamma_{\rm lm}\sum_{n=1}^{\infty}\frac{(2n-1)(n+1)}{4n+1}c_{2n}(t)\dot{c}_{2n}(t) \notag\\
&-2\pi R_{0}\gamma_{\rm lm}{\rm cos}\theta_{\rm L}\sum_{n=1}^{\infty}\dot{c}_{2n}(t)P_{2n}(0) \notag\\
&+ 2\pi \rho R_0^3 \sum_{n=1}^{\infty}\frac{\dot{c}_{2n}(t)\ddot{c}_{2n}(t)}{(2n+1)(4n+1)}.\label{EnergyChangeRate}
\end{align}
The energy dissipation function in the bulk is written as
\begin{align}
\Phi_{\rm vis}^{\rm b,u}&=\frac{\mu}{2}\int\int(\nabla \textit{v}^2)\textbf{n}~dS \notag\\
&=2\pi\mu R_0 \sum_{n=1}^{\infty}\frac{2n+2}{4n+1}\dot{c}_{2n}^2(t).\label{BulkDissipation}
\end{align}
The dissipation at the contact line is given by
\begin{align}
\Phi_{\rm vis}^{\rm ct,u}&=\pi \lambda r(\varphi=\pi/2,t) \dot{r}^2(\varphi=\pi/2,t) \notag\\
&=\pi\lambda R_0 \sum_{n=1}^{\infty}\dot{c}_{2n}^2(t)P_{2n}^2(0).\label{CTDissipation}
\end{align}
Therefore, the governing equation for the droplet shape modes is determined by the condition $\partial\Re_{\rm u}/\partial \dot{c}_{2n}(t)=0$, where $\Re_{\rm u}=\dot{E} + \Phi_{\rm vis}^{\rm b} + \Phi_{\rm vis}^{\rm ct}$ is the Rayleighian. Substituting Eqs.~(\ref{EnergyChangeRate})-(\ref{CTDissipation}) into $\partial\Re/\partial \dot{c}_{2n}(t)=0$ yields the evolution equation for the shape mode
\begin{align}
&\ddot{c}_{2n}(t) + \biggl[2\mu(2n+2) + \lambda P_{2n}^2(0)(4n+1)\biggr]\frac{2n+1}{\rho R_0^2}\dot{c}_{2n}(t)  \notag\\
&+ \frac{\gamma_{\rm lm}}{\rho R_0^3}(2n-1)(2n+1)(2n+2)c_{2n}(t) \notag\\
&=\frac{\gamma_{\rm lm}}{\rho R_0^2}(2n+1)(4n+1)P_{2n}(0){\rm cos}\theta_{\rm L}.\label{shapemode}
\end{align}
The solution to Eq.~(\ref{shapemode}) gives transient shape of EW droplets(see also Refs.~(\cite{J.M.Oh2010}) and~(\cite{Z.Sun2021})).

\section{RESULTS AND DISCUSSION}
Our calculation is carried out by using the values of parameters $\varepsilon=1.93$, $d=2.5~{\rm mm}$, $\mu_0=4.6~{\rm mPa\cdot s}$, and $C=26.24$~\cite{Q.Vo2018PRE}. The controlling parameters $U$, $\mu$, $\gamma_{\rm lm}$, and $V_0$ are given in the corresponding figures.
\subsection{Overdamped regime}
\begin{figure}[htp]
%\centerline{\includegraphics[width=1.0\textwidth,keepaspectratio]{figure2}}
  \includegraphics[width=\linewidth,keepaspectratio]{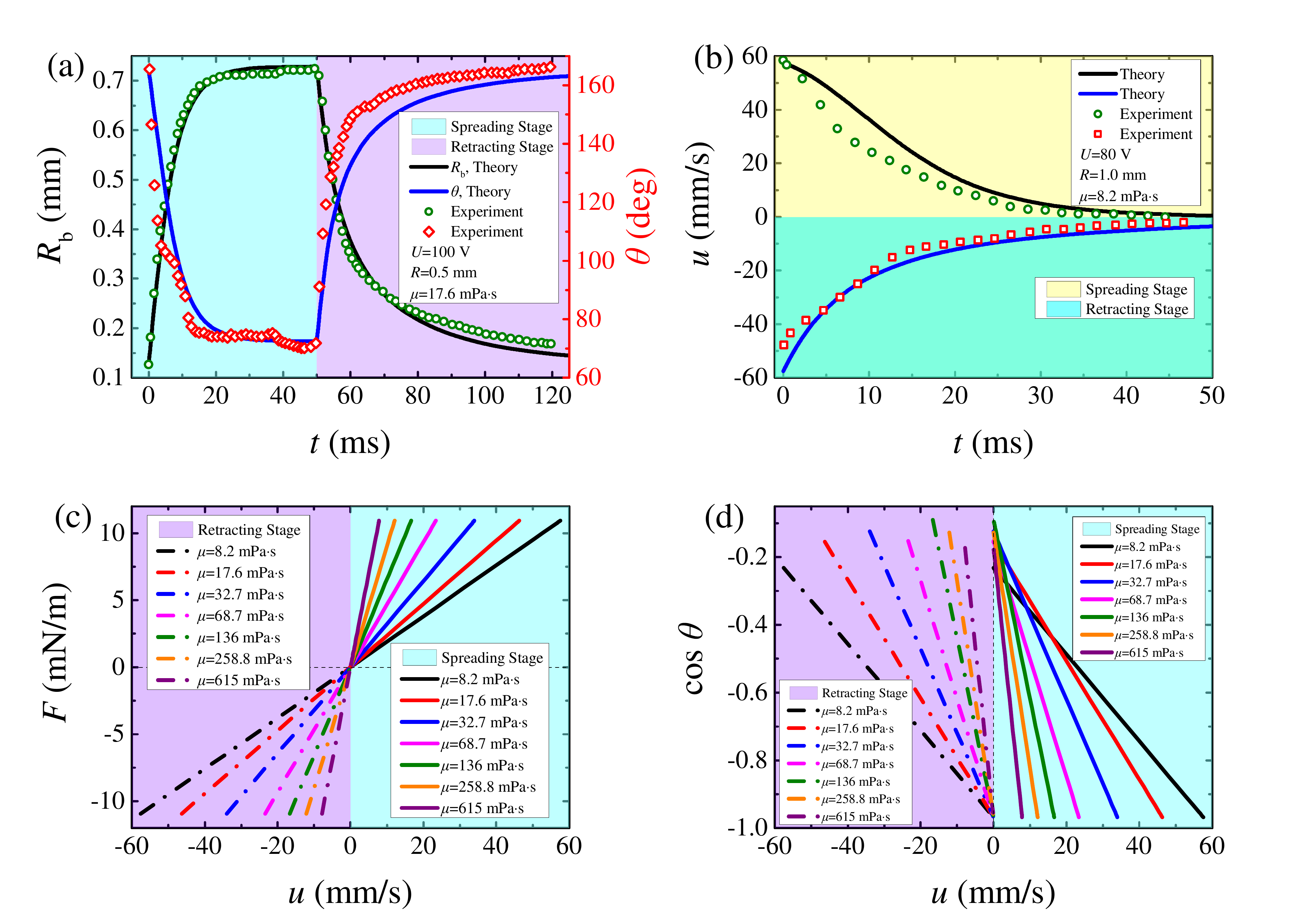}
  \caption{(Color online) Comparison of theoretical time-dependent (a) base radius $R_b$ and contact angle $\theta$, and (b) contact line velocity $u$ with those measured in Ref~\cite{Q.Vo2018PRE}. $\theta_{\rm Y}=165.4^\circ$ is used for calculation. Plots of (c) driving force $F$ and (d) ${\rm cos}~\theta$ versus contact line velocity $u$ for different droplet viscosities with $U=80~V$, initial droplet size $R=1.0~{\rm mm}$, and $\theta_{\rm Y}=165.4^\circ$. \label{RbandContactangleandVelocityVsEta}}
\end{figure}

In the overdamped regime, in order to numerically solve Eqs.~(\ref{volumeconstant}),~(\ref{SpreadingkineticEq}), and ~(\ref{RetractingkineticEq}), and compare our theoretical results with the experimental ones, we use the same values of parameters as those in Refs.~\cite{Q.Vo2018SP} and~\cite{Q.Vo2018PRE}, i.e., $\gamma_{\rm lm}=26.9$~mN~${\rm m^{-1}}$ and $\theta_{\rm Y}=165.4^\circ$. In Fig.~\ref{RbandContactangleandVelocityVsEta}(a), theoretical curves (black and blue lines) of base radius $R_{\rm b}$ and contact angle $\theta$ based on the present model are displayed to compare with the experimental results (olive circle and red diamond dots) in Ref.~\cite{Q.Vo2018PRE}.
It is shown in the figure that our theoretical prediction agree considerably well, both in the spreading and the retraction stages, with the experimental results reported by Q. Vo \textit{et al.}~\cite{Q.Vo2018PRE}, indicating that our theoretical model is a convincible one.
To further capture the dynamic features of the EW and the retraction process, we plot the three-phase contact line (TCL) velocity $u=\dot{R}_{\rm b}$ against time $t$ for these two stages, as illustrated in Fig.~\ref{RbandContactangleandVelocityVsEta}(b). It is found that the contact line velocity monotonically reduces to zero, a conclusion again in very good agreement with that obtained by Q. Vo \textit{et al.}~\cite{Q.Vo2018PRE} (circle and square symbol), where black line and circle symbol refer to spreading stage, and blue line and square symbol corresponds to retraction process respectively. The monotonic variation trend of $R_{\rm b}$, $\theta$, and $u$ demonstrates that the contact line, in both spreading and retraction stages, evolves in an overdamped way. The reason lies in that the liquid inertia effect can be neglected and viscosity becomes a dominant factor causing the droplet to spread or retract gradually to its equilibrium state. In particular, contact line friction dissipates most excessive interfacial energy in the overdamped regime.

It is widely known that there exists a driving force per unit length that pulls the liquid at the three-phase contact line~\cite{F.Mugele2005,K.H.Kang2002}, which is given by $F=\gamma_{\rm lm}({\rm cos}~\theta_{\rm L}-{\rm cos}~\theta)$ for the spreading stage ($F=\gamma_{\rm lm}({\rm cos}~\theta_{\rm L}-{\rm cos}~\theta)$ for retraction stage) according to the interfacial tension imbalance in the horizontal direction. Figure~\ref{RbandContactangleandVelocityVsEta}(c) plots the dependence of $F$ on $u$ for various droplet viscosities. The linear dependence between $F$ and $u$ suggests a linear relation $F=\lambda u$, which is also in line with the previous reports that the balance between the driving force and the contact line friction force yields a linear relation between $F$ and $u$~\cite{Q.Vo2018PRE,Q.Vo2018SP,Q.G.Wang2020,K.Xiao2021}.
Here, $\lambda$ is the friction coefficient for the three-phase contact line (TCL), which originates from the interactions between the liquid molecules and the solid surface
at TCL~\cite{T.D.Blake1969}.

The dependence of the dynamic contact angle, $\theta$, on the speed of the contact line, $u$, is a critical consideration in wetting dynamics. To explore the relation between them, the dependence of ${\rm cos}~\theta$ on $u$ for different droplet viscosities is plotted in Fig.~\ref{RbandContactangleandVelocityVsEta}(d). It is found that ${\rm cos}~\theta$ exhibits a linear relation with $u$ in both EW and retraction, which is in consistence with the previous reports as well~\cite{H.Li2013}.
The quantitatively good agreement between theoretical and experimental results verify that the droplet spreading and the droplet retraction dynamic behaviors driven by EW can be analyzed very well by our model in the overdamped regime (i.e., droplet with high viscosity).

\begin{figure}[htp]
  \includegraphics[width=\linewidth,keepaspectratio]{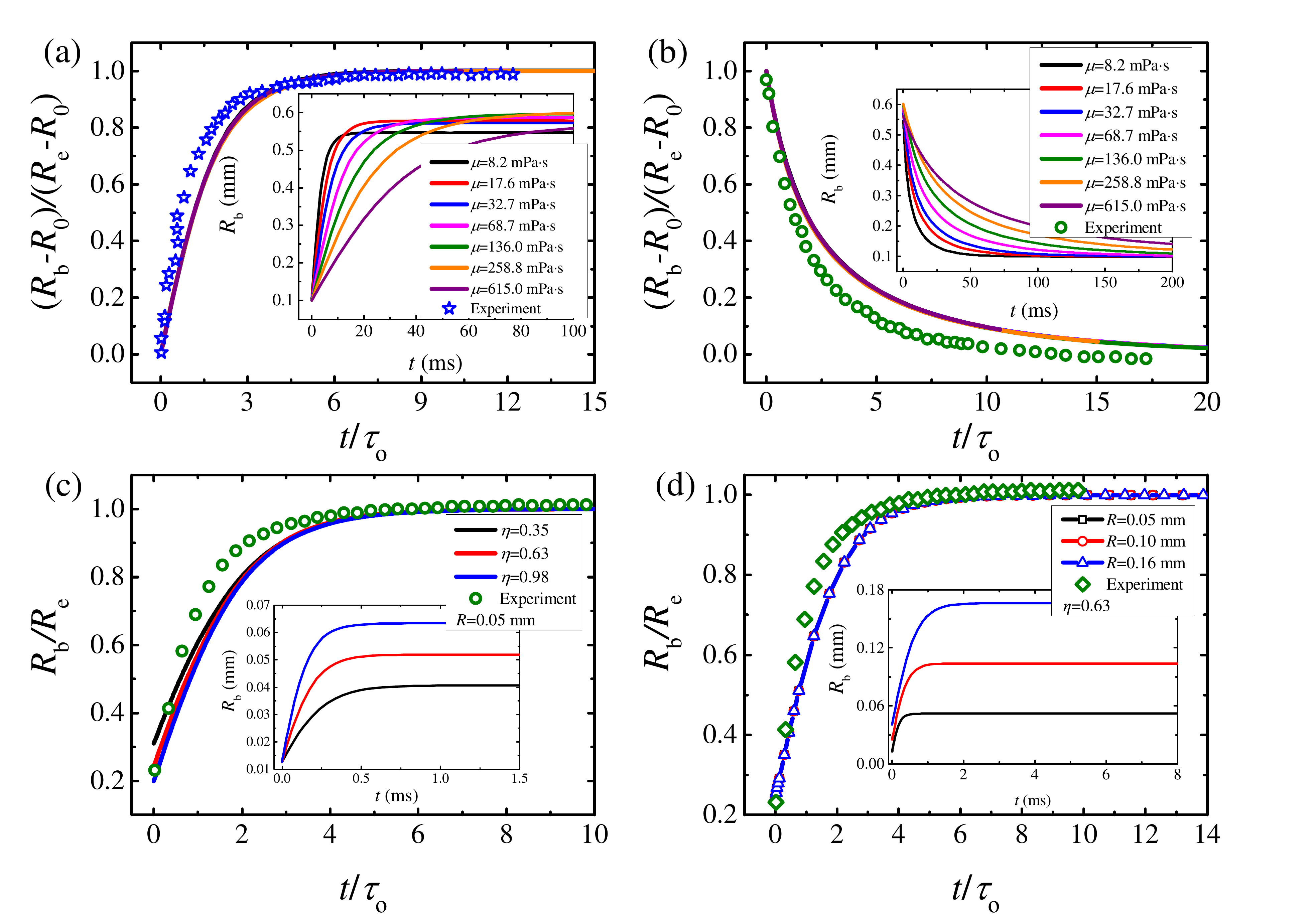}
  \caption{(Color online) Plots of rescaled base radius $(R_{\rm b}-R_0)/(R_{\rm e}-R_0)$ versus normalized time $t/\tau_{\rm o}$ for (a) spreading stage and (b) recoiling stage, with parameter values $U=80~V$ and $R=1~{\rm mm}$. Plots of normalized spreading base radius $R_{\rm b}/R_{\rm e}$ versus normalized time $t/\tau_{\rm o}$ for different (c) EW numbers $\eta$  and (d) droplet sizes $R$. The experimental data in (a) and (b), and (c) and (d) come from Ref.~\cite{Q.Vo2018PRE} and Ref.~\cite{Q.Vo2018SP} respectively.\label{RdVsTo}}
\end{figure}

Here it should be noted that in the overdamped regime, the viscous dissipation in the bulk is assumed to be negligible due to the small velocity gradient. In fact, there exists a characteristic timescales $\tau_{\rm o}$ for the transient dynamics of viscous droplets actuated by EW. The balance between the driving force and the contact line friction force yields a characteristic timescale $\tau_{\rm o}=\lambda (R_{\rm e}-R_0)/(\eta\gamma_{\rm lm})$, which represents the timescale for a droplet to switch between the initial and the final equilibrium states.
To confirm that $\tau_{\rm o}$ is the characteristic timescale for the transient dynamics of droplets spreading and retraction in ambient environment, we first respectively exhibit the temporal evolution of base radius for various droplet viscosities during the EW and retraction processes in Figs.~\ref{RdVsTo}(a) and (b), where the insets are the corresponding raw results of each plot. Here, the time of evolution is rescaled as $t/\tau_{\rm o}$, while the base radius is rescaled as $(R_{\rm b}-R_0)/(R_{\rm e}-R_0)$. Notably, the rescaled base radius follows a master curve, demonstrating that normalized time-dependent base radius is independent of liquid viscosities. In contrast, the insets in Figs.~\ref{RdVsTo}(a) and (b) illustrate the effects of liquid properties.

Figures~\ref{RdVsTo}(c) and (d) show the effects of applied voltage and droplet size on spreading base radius. Here the base radius is normalized by the base radius of final equilibrium state, and insets demonstrate plots of $R_{\rm b}$ versus $t$ prior to normalization. Similarly, the normalized evolutions collapse to a master curve independent of applied voltage and droplet size, indicating that timescale $\tau_{\rm o}$ can be used to characterize the transient dynamics in the overdamped regime. This can be ascribed to the fact that the viscous dissipation in the bulk has been neglected. In addition, Fig.~\ref{RdVsTo} also compares our theoretical results with the experimental data from Refs.~\cite{Q.Vo2018SP} and~\cite{Q.Vo2018PRE}, the good agreement between them verifies that the present theoretical model based on Onsager variational principle is suitable to interpret non-equilibrium dynamics in EW-related problems.

\subsection{Underdamped regime}
\begin{figure}[htp]
  \includegraphics[width=\linewidth,keepaspectratio]{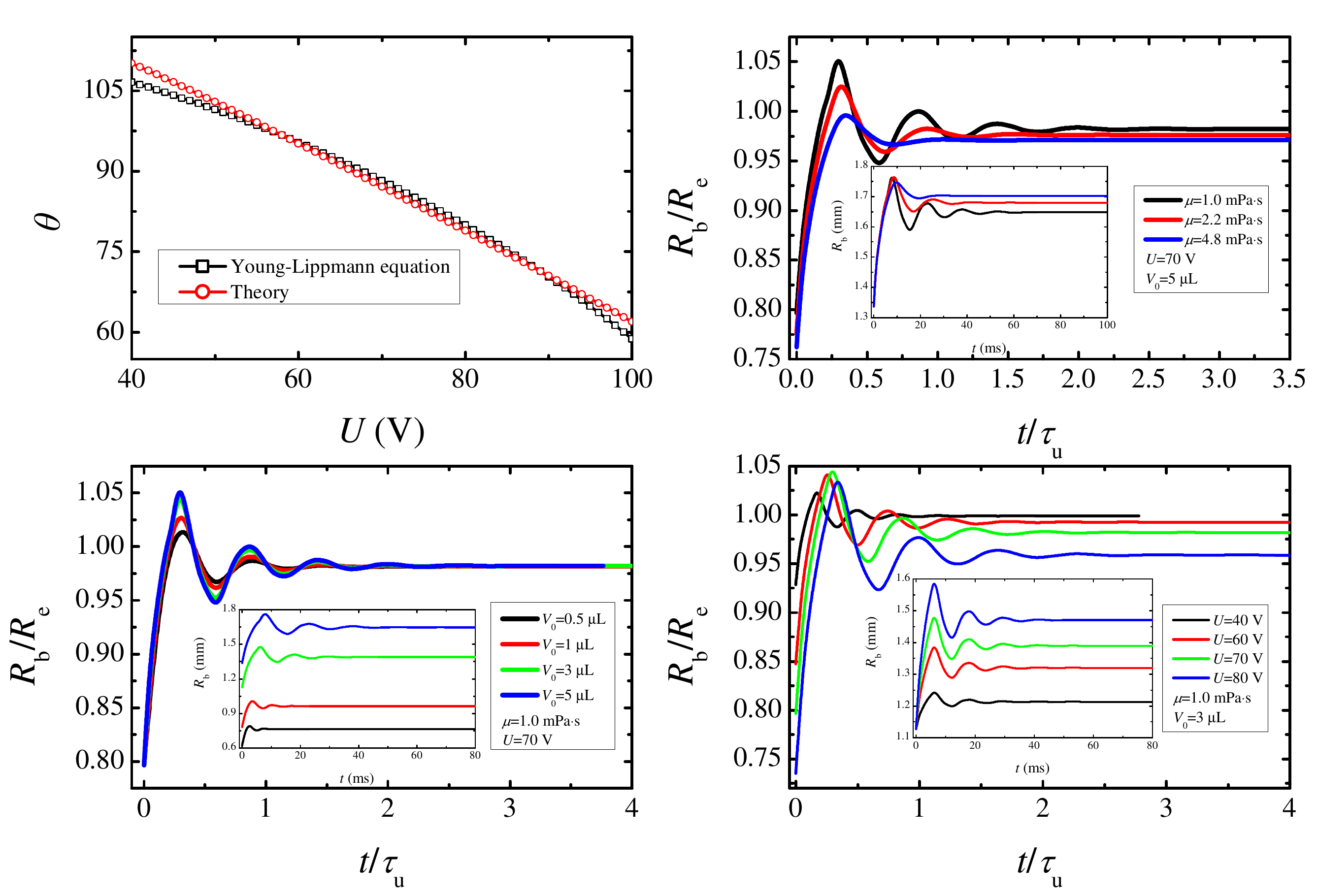}
  \caption{(Color online) (a) Comparison of contact angle predicted by the present model with that calculated by Young-Lippmann equation. Plots of normalized spreading base radius $R_{\rm b}/R_{\rm e}$ versus normalized time $t/\tau_{\rm u}$ for different (b) droplet viscosities $\mu$, (c) applied voltages $U$, and (d) droplet volumes $V_0$.\label{RdVsTu}}
\end{figure}

Besides overdamped regime, we also investigate the spreading dynamics of droplets with low viscosity, namely the underdamped regime. Our calculation is performed by using the values of parameters $\varepsilon=1.93$, $d=2.5~{\rm mm}$, $\mu_0=4.6~{\rm mPa\cdot s}$, and $C=32.9$~\cite{Q.Vo2018SP}. All the summations in the paper were carried out over $n=120$. In the underdamped regime, our analysis of EW dynamics is based on the derived shape equation Eq.~(\ref{shapemode}), thus it is necessary to verify the validity of the present model. For this reason, we examine whether the apparent contact angle at equilibrium state predicted by our model is consistent with the well-known Young-Lippmann equation~\cite{G.Lippmann1875,F.Mugele2005} ${\rm cos}~\theta_{\rm L}={\rm cos}~\theta_{\rm Y}+\varepsilon_0 \varepsilon U^2/(2d\gamma_{\rm lm})$. We numerically calculate the apparent contact angle by utilizing Eq.~(27) in Ref.~\cite{J.M.Oh2010}, and the comparison between the contact angle predicted by the present model and those of Young-Lippmann equation is exhibited in Fig.~\ref{RdVsTu}(a).
It is observable that the calculated contact angle based on the present model agrees very well with that of Young-Lippmann equation, an evidence once again confirming the validity of the present model.

Subsequently, in order to reveal how droplet viscosity, droplet size, and applied voltage affect the spreading dynamics driven by EW, we plot the normalized spreading base radius $R_{\rm b}/R_{\rm e}$ versus normalized time $t/\tau_{\rm u}$ for various controlling factors (liquid viscosity, droplet size, and applied voltage), as demonstrated in Fig.~\ref{RdVsTu}(b), (c) and (d). The insets show the corresponding unnormalized $R_{\rm b}$. In the figure it is found that a smaller droplet with higher viscosity under lower voltage activation reaches its maximum peak and equilibrium shape faster than a larger droplet with lower viscosity. Here, $\tau_{\rm u}$ is the characteristic timescale for the droplet to reach
maximum deformation in the underdamped regime, which is given by $\tau_{\rm u}=\pi\rho^{1/2}R^{3/2}/(\eta\gamma_{\rm lm})^{1/2}$ based on the balance between the driving force and droplet's inertia with the viscosity neglected~\cite{Q.Vo2018SP,Q.Vo2019}. The time evolution of the normalized spreading base radius in Fig.~\ref{RdVsTu} indicates that $R_{\rm b}$ monotonically increases to the first peak and reaches the maximum deformation in the beginning, followed by a damped oscillation, and eventually approaches a stable equilibrium state.
The apparent oscillation at the droplet's surface stems from the from the fact that the droplet's inertia resists the contact line motion even though eventually the corresponding energy is absorbed by the contact line as a result of dissipation.
 The conclusion that $\tau_{\rm u}$ is a characteristic timescale in the underdamped regime does not change for different controlling parameters, such as droplet viscosity (Fig.~\ref{RdVsTu}(b)), droplet size (Fig.~\ref{RdVsTu}(c)), and applied voltage (Fig.~\ref{RdVsTu}(d)).
Our theoretical predictions also agree well with the experimental results reported in Refs.~\cite{Q.Vo2018SP,J.M.Oh2010,J.Hong2013}.

Consequently, the good agreement between theoretical prediction and experimental results suggests that both the spreading and the retraction dynamics driven by EW can be captured by our proposed theoretical model. In the overdamped regime, the damping effect due to viscous dissipation and friction dominates the whole spreading or the retraction process, which makes the droplet to spread or retract gradually to its equilibrium configuration without oscillation. In this regime, the transient dynamics can be characterized by a typical timescale $\tau_{\rm o}$.
Whereas in the underdamped regime, the inertial effect is comparable with the damping effect, resulting in an apparent oscillation
on the droplet's surface. Similarly, its transient dynamics can be characterized by a timescale $\tau_{\rm u}$.
It is found that the characteristic timescales $\tau_{\rm o}$ and $\tau_{\rm u}$ are independent of liquid viscosity, droplet size and applied voltage.

\section{CONCLUSION}
In summary, we develop a theoretical model for the EW dynamics of a viscous droplet immersed in an ambient environment by using Onsager principle, and obtain an array of equations governing the EW-driven spreading dynamics. The excellent agreement between our numerical results and the corresponding experimental ones shows that our model captures the essential phenomena of the dynamic behaviors of both spreading and retraction driven by EW. Our theoretical results also reveal that the transient dynamics of viscous droplets can be characterized by a typical timescale $\tau_{\rm o}$ or $\tau_{\rm u}$, respectively, for the overdamped or the underdamped regime. Such characteristic timescales are independent of liquid viscosity, droplet size and applied voltage.

\begin{acknowledgements}
We acknowledge the financial supports from National Natural Science Foundation of China under Grant Nos.12147142, 11974292, 12174323, and 1200040838. Thanks are also given to Xi Chen, Wei Li, and Dr. Rui Ma for helpful discussions.
\end{acknowledgements}

\section{Appendix: }

%\bibliography{swimmer} %You need to replace "rsc" on this line with the name of your .bib file

\end{document}